\documentclass[aps,prd,onecolumn,groupedaddress,showpacs,nofootinbib,amssymb]{revtex4}
\usepackage[english]{babel}
\usepackage{bm}
\usepackage{amsmath,amssymb,amsfonts}

\usepackage{ifpdf}
\ifpdf
  \usepackage[pdftex]{hyperref}
\else
  \usepackage[dvips]{hyperref}
\fi

\begin{document}

\newcommand{\be}{\begin{equation}}
\newcommand{\ee}{\end{equation}}
\newcommand{\bea}{\begin{eqnarray}}
\newcommand{\eea}{\end{eqnarray}}
\newcommand{\nn}{\nonumber \\}
\newcommand{\p}{\partial}

\newcommand{\mc}[1]{\mathcal{#1}}
\newcommand{\mr}[1]{\mathrm{#1}}
\newcommand{\rd}{\mathrm{d}}
\newcommand{\vect}[1]{\bm{#1}}

\newcommand{\g}[1][]{g^{(3)#1}}
\newcommand{\dg}[1][]{\dot{g}^{(3)#1}}

\title{Modified $F(R)$ Ho\v{r}ava-Lifshitz gravity: a way to accelerating FRW 
cosmology}

\author{Masud Chaichian$^{1,2}$,
Shin'ichi Nojiri$^3$, Sergei D. Odintsov$^{4,5}$\footnote{Also
at Tomsk State Pedagogical University},
Markku Oksanen$^1$, and Anca Tureanu$^{1,2}$}
\affiliation{
$^1$ Department of Physics, University of Helsinki, P.O. Box 64, FI-00014 Helsinki, Finland \\
$^2$ Helsinki Institute of Physics, P.O. Box 64, FI-00014 Helsinki, Finland
\\
$^3$ Department of Physics, Nagoya University, Nagoya 464-8602, Japan \\
$^4$ Instituci\`{o} Catalana de Recerca i Estudis Avan\c{c}ats (ICREA),
Barcelona \\
$^5$ Institut de Ciencies de l'Espai (IEEC-CSIC),
Campus UAB, Facultat de Ciencies, Torre C5-Par-2a pl, E-08193 Bellaterra
(Barcelona), Spain}

\begin{abstract}

We propose a general approach for the construction of modified gravity 
which is invariant under foliation-preserving diffeomorphisms. Special 
attention is paid to the formulation of modified $F(R)$ 
Ho\v{r}ava-Lifshitz gravity (FRHL), whose Hamiltonian structure is 
studied. It is demonstrated that the spatially-flat FRW equations of 
FRHL are consistent with the constraint equations. The analysis of 
de Sitter solutions for several versions of FRHL indicates that the 
unification of the early-time inflation with the late-time acceleration 
is possible. It is shown that a special choice of parameters for FRHL 
leads to the same spatially-flat FRW equations as in the case of 
traditional $F(R)$-gravity. Finally, an essentially most general 
modified Ho\v{r}ava-Lifshitz gravity is proposed, motivated by its 
fully diffeomorphism-invariant counterpart, with the restriction that 
the action does not contain derivatives higher than the second 
order with respect to the time coordinate.

\end{abstract}


\maketitle

\section{Introduction}

Recent observational data clearly indicates that our universe is
currently expanding with an accelerating rate, apparently due to
Dark Energy. The early universe has also undergone a period of
accelerated expansion (inflation). The modified gravity approach
(for a general review, see \cite{review}) suggests that such
accelerated expansion is caused by a modification of gravity at the
early/late-time universe. A number of modified theories of gravity,
which successfully describe the unification of early-time inflation
with late-time acceleration and which are cosmologically and
observationally viable, has been proposed (for a review, see
\cite{review}). Despite some indications \cite{M} that such
alternative theories of gravity may emerge from string/M-theory,
they are still mostly phenomenological theories that are not yet
related to a fundamental theory.

Recently the so-called Ho\v{r}ava-Lifshitz quantum gravity
\cite{Horava:2009uw} has been proposed. This theory appears to be
power-counting renormalizable in 3+1 dimensions. One of the key
elements of such a formulation is to abandon the local Lorentz
invariance so that it is restored as an approximate symmetry at low
energies. Despite its partial success as a candidate for a
fundamental theory of gravity, there are a number of unresolved
problems (see refs. \cite{Charmousis:2009,Li:2009,Blas:2009:1,
Kobakhidze:2009,Koyama:2009,Henneaux:2009zb})
related with the detailed balance and the projectability conditions
(see section \ref{sec2} for definitions), strong couplings,
an extra propagating degree of freedom and the GR (infrared) limit,
the relation with other modified theories of gravity etc.
Moreover, study of the spatially-flat FRW cosmology
in the Ho\v{r}ava-Lifshitz gravity indicates that its background
cosmology \cite{cosmology} is almost the same as in the usual 
GR, although an effective dark matter could appear as a kind of 
a constant of integration in the Ho\v{r}ava-Lifshitz gravity 
\cite{Mukohyama:2009mz}. 
Hence, it seems that there is no natural way (without extra fields)
to obtain an accelerating universe from Ho\v{r}ava-Lifshitz gravity,
let alone a unified description of the early-time inflation with the
late-time acceleration. Therefore it is natural to search for a
generalization of the Ho\v{r}ava-Lifshitz theory that could be
easily related to a traditional modified theory of gravity. On the
one hand, it may be very useful for the study of the low-energy
limit of such a generalized Ho\v{r}ava-Lifshitz theory due to the
fact that a number of modified theories of gravity are
cosmologically viable and pass the local tests. On the other hand,
it is expected that such a generalized Ho\v{r}ava-Lifshitz gravity
may have a much richer cosmological structure, including the
possibility of a unification of the early-time inflation with the
late-time acceleration. Finally, within a more general theory one
may hope to formulate the dynamical scenario for the Lorentz
symmetry violation/restoration caused by the expansion of the
universe.

In the present work we propose such a general modified
Ho\v{r}ava-Lifshitz gravity. We mainly consider modified $F(R)$
Ho\v{r}ava-Lifshitz gravity which is shown to coincide with the
traditional $F(R)$-gravity on the spatially-flat FRW background for
a special choice of parameters. Another limit of our model leads to
the degenerate $F(R)$ Ho\v{r}ava-Lifshitz gravity proposed in ref.
\cite{Kluson:2009xx}. The Hamiltonian analysis of the modified
$F(R)$ Ho\v{r}ava-Lifshitz theory is presented. The preliminary
investigation of the FRW equations for models from this class
indicates a rich cosmological structure and a natural possibility
for the unification of the early-time inflation with the Dark Energy
epoch. Finally, we propose the most general modification of
Ho\v{r}ava-Lifshitz-like theory of gravity. Our formulation ensures
that the spatially-flat FRW cosmology of any modified
Ho\v{r}ava-Lifshitz gravity (for a special choice of parameters)
coincides with the one of its traditional modified gravity
counterpart.

\section{Modified $F(R)$ Ho\v{r}ava-Lifshitz gravity}
\label{sec2}

In this section we propose a new extended action for $F(R)$
Ho\v{r}ava-Lifshitz gravity. The FRW equations for this theory are
also formulated. The action of the standard $F(R)$-gravity is given
by
\be
\label{HLF1} S_{F(R)} = \int \rd^4 x \sqrt{-g} F(R)\, .
\ee
Here $F$ is a function of the scalar curvature $R$. By using the ADM
decomposition \cite{Arnowitt:1962hi} (for reviews and mathematical
background see \cite{Wald:1984,Gourgoulhon:2007}), we can write the metric in 
the
following form:
\be
\label{HLF2}
ds^2 = - N^2 \rd t^2 +
g^{(3)}_{ij}\left(\rd x^i + N^i \rd t \right)\left(\rd x^j + N^j \rd
t \right), \quad i=1,2,3\, .
\ee
Here $N$ is called the lapse
variable and $N^i$'s are the shift variables. Then the scalar
curvature $R$ has the following form:
\be
\label{HLF3} R= K^{ij}
K_{ij} - K^2 + R^{(3)} + 2 \nabla_\mu \left( n^\mu \nabla_\nu n^\nu
  - n^\nu \nabla_\nu n^\mu \right) \,
\ee
and $\sqrt{-g} =
\sqrt{g^{(3)}} N$. Here $R^{(3)}$ is the three-dimensional scalar
curvature defined by the metric $g^{(3)}_{ij}$ and $K_{ij}$ is the
extrinsic curvature defined by
\be
\label{HLF4}
K_{ij}=\frac{1}{2N}\left(\dot
g^{(3)}_{ij}-\nabla^{(3)}_iN_j-\nabla^{(3)}_jN_i\right) \, ,\quad K
=K^i_{\ i}\, .
\ee
$n^\mu$ is a unit vector perpendicular to the
three-dimensional hypersurface $\Sigma_t$ defined by $t=\text{constant}$
and $\nabla^{(3)}_i$ expresses the covariant derivative on
the hypersurface $\Sigma_t$.

\newcommand{\lambdaw}{\Lambda_W^{}}
\newcommand{\bx}{\bm{x}}
\newcommand{\CG}{\mathcal{G}}
\newcommand{\trace}{\mathrm{Tr}}

Recently an extension of $F(R)$-gravity to a Ho\v{r}ava-Lifshitz
type theory \cite{Horava:2009uw} has been proposed
\cite{Kluson:2009xx}, by introducing the action
\be
\label{HLF5}
S_{F_\mathrm{HL}(R)} = \int \rd^4 x \sqrt{g^{(3)}} N
F(R_\mathrm{HL})\, ,\quad R_\mathrm{HL} \equiv K^{ij} K_{ij} -
\lambda K^2 - E^{ij}\CG_{ijkl} E^{kl} \, . \ee Here $\lambda$
is a real constant in the ``generalized De~Witt metric'' or
``super-metric'' (``metric of the space of metric''),
\be
\label{HLF6}
\CG^{ijkl} = \frac{1}{2}\left( g^{(3) ik} g^{(3) jl} + g^{(3) il} g^{(3) jk} 
\right)
- \lambda g^{(3) ij} g^{(3) kl}\, ,
\ee
defined on the three-dimensional hypersurface $\Sigma_t$,
$E^{ij}$ can be defined by the so called \emph{detailed balance condition}
by using an action $W[g^{(3)}_{kl}]$ on the hypersurface $\Sigma_t$
\be
\label{HLF7}
\sqrt{g^{(3)}}E^{ij}=\frac{\delta W[g^{(3)}_{kl}]}{\delta g_{ij}}\, ,
\ee
and the inverse of $\CG^{ijkl}$ is written as
\be
\CG_{ijkl} = \frac{1}{2}\left( g^{(3)}_{ik} g^{(3)}_{jl}
+ g^{(3)}_{il} g^{(3)}_{jk} \right)
  - \tilde{\lambda} g^{(3)}_{ij} g^{(3)}_{kl}\, ,\quad
\tilde{\lambda} = \frac{\lambda}{3\lambda - 1}\, .
\ee
The action $W[g^{(3)}_{kl}]$ is assumed to be defined by the metric and
the covariant derivatives on the hypersurface $\Sigma_t$.
The original motivation for the detailed balance condition is its ability to
simplify the quantum behaviour and renormalization properties
of theories that respect it. Otherwise there is no a priori physical
reason to restrict $E^{ij}$ to be defined by (\ref{HLF7}).
There is an anisotropy between space and time in the Ho\v{r}ava-Lifshitz
gravity. In the ultraviolet (high energy) region, the time
coordinate and the spatial coordinates are assumed to behave as
\be
\label{HLF7b}
\bm{x}\to b\bm{x}\ ,\quad t\to b^z t\ ,\quad
z=2,3,\cdots\, ,
\ee
under the scale transformation. In
\cite{Horava:2009uw}, $W[g^{(3)}_{kl}]$ is explicitly given for the
case $z=2$,
\be
\label{HLF7c} W=\frac{1}{\kappa_W^2}\int
\rd^3\vect{x}\,\sqrt{g^{(3)}}(R-2\lambdaw)\, ,
\ee
and for the case $z=3$,
\be
\label{HLF7d}
W=\frac{1}{w^2}\int_{\Sigma_t}\omega_3(\Gamma)\, .
\ee
Here $\kappa_W$
in (\ref{HLF7c}) is a coupling constant of dimension $-1/2$ and
$w^2$ in (\ref{HLF7d}) is the dimensionless coupling constant.
$\omega_3(\Gamma)$ in (\ref{HLF7d}) is given by
\be
\label{HLF7e}
\omega_3(\Gamma) = \trace\left(\Gamma\wedge
d\Gamma+\frac{2}{3}\Gamma\wedge\Gamma \wedge\Gamma\right) \equiv
\varepsilon^{ijk}\left(\Gamma^{m}_{il}\p_j
\Gamma^{l}_{km}+\frac{2}{3}\Gamma^{n}_{il}\Gamma^{l}_{jm}
\Gamma^{m}_{kn}\right)\rd^3\vect{x}\, .
\ee
A general $E^{ij}$ consist of all contributions to $W$ up to the chosen value 
$z$.

In the Ho\v{r}ava-Lifshitz-like $F(R)$-gravity, we assume that $N$
can only depend on the time coordinate $t$, which is called the
\emph{projectability condition}. The reason is that the
Ho\v{r}ava-Lifshitz gravity does not have the full diffeomorphism
invariance, but is invariant only under ``foliation-preserving''
diffeomorphisms, i.e. under the transformations
\be
\label{fpd1}
\delta x^i=\zeta^i(t,\bm{x})\,, \, \quad \delta t=f(t)\, .
\ee
If $N$ depended on the spatial coordinates, we could not fix $N$ to be
unity ($N=1$) by using the foliation-preserving diffeomorphisms.
There exists a version of Ho\v{r}ava-Lifshitz gravity without
the projectability condition, but it is suspected to possess few
additional consistency problems \cite{Henneaux:2009zb,Li:2009}.
Therefore we prefer to assume that $N$ depends only on the time
coordinate $t$.

Let us consider the FRW universe with a flat spatial part,
\be
\label{HLF8}
ds^2 = - N^2 \rd t^2 + a(t)^2 \sum_{i=1,2,3} \left( \rd x^i \right)^2\, .
\ee
Then, it is clear from the explicit
expressions in (\ref{HLF7c}) and (\ref{HLF7d}) that $W[g^{(3)}_{kl}]$
vanishes identically if $\lambdaw =0$, which we assume since
a non-vanishing $\lambdaw$ gives a cosmological constant. Then one can
obtain
\be
\label{HLF9}
R= \frac{ 12 H^2}{N^2} +
\frac{6}{N}\frac{\rd}{\rd t}\left(\frac{H}{N}\right) = -
\frac{6H^2}{N} + \frac{6}{a^3 N} \frac{\rd}{\rd t}\left(
\frac{Ha^3}{N} \right)\, ,\quad R_\mathrm{HL} = \frac{ \left(3 - 9
\lambda \right) H^2}{N^2} \, .
\ee
Here the Hubble rate $H$ is
defined by $H\equiv \dot a/ a$. In the case of the Einstein gravity,
the second term in the last expression for $R$ becomes a total
derivative:
\be
\label{HLF10}
\int \rd^4 x \sqrt{-g} R = \int \rd^4 x\ a^3 N
\left\{ - \frac{6H^2}{N} + \frac{6}{a^3 N} \frac{\rd}{\rd t}
\left( \frac{Ha^3}{N} \right) \right\}
= \int \rd^4 x \left\{ - 6H^2 a^3 + 6 \frac{\rd}{\rd t}
\left( \frac{Ha^3}{N} \right) \right\}
\, .
\ee
Therefore, this term can be dropped in the Einstein
gravity. The total derivative term comes from the last term
$2\nabla_\mu \left( n^\mu \nabla_\nu n^\nu - n^\nu \nabla_\nu n^\mu
\right)$ in (\ref{HLF3}), which is dropped in the usual
Ho\v{r}ava-Lifshitz gravity. In the $F(R)$-gravity, however, this
term cannot be dropped due to the non-linearity. Then if we consider
the FRW cosmology with the flat spatial part, there is almost no
qualitative difference between the Einstein gravity and the
Ho\v{r}ava-Lifshitz gravity, except that there could appear an
effective dark matter as a kind of a constant of integration in the
Ho\v{r}ava-Lifshitz gravity \cite{Mukohyama:2009mz}. The effective
dark matter appears since the constraint given by the variation over
$N$ becomes global in the projectable Ho\v{r}ava-Lifshitz gravity.

Now we propose a new and very general Ho\v{r}ava-Lifshitz-like
$F(R)$-gravity by
\be
\label{HLF11}
S_{F(\tilde R)} = \int \rd^4 x \sqrt{g^{(3)}} N F(\tilde R)\, , \quad
\tilde R \equiv K^{ij} K_{ij} - \lambda K^2
+ 2 \mu \nabla_\mu \left( n^\mu \nabla_\nu n^\nu - n^\nu \nabla_\nu
n^\mu \right) - E^{ij}\CG_{ijkl} E^{kl} \, .
\ee
In the FRW
universe with the flat spatial part, $\tilde R$ has the following
form:
\be
\label{HLF12}
\tilde R= \frac{ \left(3 - 9 \lambda \right)
H^2}{N^2} + \frac{6\mu }{a^3 N} \frac{\rd}{\rd t}\left(
\frac{Ha^3}{N} \right) = \frac{ \left(3 - 9 \lambda + 18 \mu \right)
H^2}{N^2} + \frac{6\mu }{N} \frac{\rd}{\rd t}\left( \frac{H}{N}
\right)\, .
\ee
The case one obtains with the choice of parameters
$\lambda=\mu=1$ corresponds to the usual $F(R)$-gravity as long as
we consider spatially-flat FRW cosmology, since $\tilde R$ reduces
to $R$ in (\ref{HLF9}). On the other hand, in the case of $\mu=0$,
$\tilde R$ reduces to $R_\mathrm{HL}$ in (\ref{HLF9}) and therefore
the action (\ref{HLF11}) becomes identical with the action
(\ref{HLF5}) of the Ho\v{r}ava-Lifshitz-like $F(R)$-gravity in
\cite{Kluson:2009xx}. Hence, the $\mu=0$ version corresponds to some
degenerate limit of the above general $F(R)$ Ho\v{r}ava-Lifshitz
gravity. We call this limit degenerate because it is very 
difficult (perhaps even impossible) to obtain FRW equations when 
$\mu=0$ is set from the very begining. In our theory the FRW equations 
can be obtained quite easily, and then $\mu=0$ is a simple limit. 

For the action (\ref{HLF11}), the FRW equation given by the variation
over $g^{(3)}_{ij}$ has the following form after assuming
the FRW space-time (\ref{HLF8}) and setting $N=1$:
\be
\label{HLF13}
0 = F\left(\tilde R\right) - 2 \left(1 - 3\lambda + 3\mu \right)
\left(\dot H + 3 H^2\right)
F'\left(\tilde R\right) - 2\left(1 - 3\lambda \right) H \frac{d
F'\left(\tilde R\right)}{\rd t}
+ 2\mu \frac{\rd^2 F'\left(\tilde R\right)}{\rd t^2} + p\, ,
\ee
where $F'$ denotes the derivative of $F$ with respect to its argument.
Here, the matter contribution (the pressure $p$) is included.
On the other hand, the variation over $N$ gives the global constraint:
\be
\label{HLF14}
0 = \int \rd^3 \vect{x} \left[ F\left(\tilde R\right)
  - 6 \left\{ \left(1 - 3\lambda + 3\mu\right) H^2 + \mu \dot H
\right\} F'\left(\tilde R\right) + 6 \mu H \frac{\rd F'\left(\tilde
R\right)}{\rd t} - \rho \right]\, ,
\ee
after setting $N=1$.
Here $\rho$ is the energy density of matter.
Since $N$ only depends on
$t$, but does not depend on the spatial coordinates, we only obtain
the global constraint given by the integration. If the standard
conservation law is used,
\be
\label{HLF15} 0= \dot \rho + 3H
\left(\rho + p\right)\, ,
\ee
Eq. (\ref{HLF13}) can be integrated to give
\be
\label{HLF16}
0 = F\left(\tilde R\right)
  - 6 \left\{ \left(1 - 3\lambda + 3\mu\right) H^2 + \mu \dot H
\right\} F'\left(\tilde R\right) + 6 \mu H \frac{\rd F'\left(\tilde
R\right)}{\rd t} - \rho - \frac{C}{a^3}\, .
\ee
Here $C$ is the
integration constant. Using (\ref{HLF14}), one finds $C=0$. In
\cite{Mukohyama:2009mz}, however, it has been claimed that $C$ need
not always vanish in a local region, since (\ref{HLF14}) needs to be
satisfied in the whole universe. In the region $C>0$, the $Ca^{-3}$
term in (\ref{HLF16}) may be regarded as dark matter.

Note that Eq. (\ref{HLF16}) corresponds to the first FRW equation
and (\ref{HLF13}) to the second one. Specifically, if we choose
$\lambda=\mu=1$ and $C=0$, Eq. (\ref{HLF16}) reduces to
\bea
\label{HLF17}
0 &=& F\left(\tilde R\right)
  - 6 \left(H^2 + \dot H \right) F'\left(\tilde R\right)
+ 6 H \frac{\rd F'\left(\tilde R\right)}{\rd t} - \rho \nn
&=& F\left(\tilde R\right)
  - 6 \left(H^2 + \dot H \right) F'\left(\tilde R\right)
+ 36 \left(4H^2 \dot H + \ddot H\right) F''\left(\tilde R\right)
  - \rho \, ,
\eea
which is identical to the corresponding equation in
the standard $F(R)$-gravity (see Eq. (2) in \cite{Nojiri:2009kx}
where a reconstruction of the theory has been made).

We should note that in the degenerate $\mu=0$ case
\cite{Kluson:2009xx}, the action (\ref{HLF11}) or (\ref{HLF5}) does
not contain any term with second derivatives with respect to the
coordinates, which appears in the usual $F(R)$-gravity. The
existence of the second derivatives in the usual $F(R)$-gravity
induces the third and fourth derivatives in the FRW equation as in
(\ref{HLF13}). Due to such higher derivatives, there appears an
extra scalar mode, which is often called the scalaron in the usual
$F(R)$-gravity. This scalar mode often affects the correction to the
Newton law as well as other solar tests. Therefore, such a scalar
mode does not appear in the $F(R)$ Ho\v{r}ava-Lifshitz gravity
with $\mu=0$. Hence, we have formulated a general
Ho\v{r}ava-Lifshitz $F(R)$-gravity which describes the standard
$F(R)$-gravity or its non-degenerate Ho\v{r}ava-Lifshitz extension
in a consistent way.

\section{Hamiltonian formalism}

Let us present some elements of the Hamiltonian analysis of our
proposal (for Hamiltonian analysis of constrained systems, and
their quantization, see \cite{HamiltonianAnalysis}). By introducing
two auxiliary fields $A$ and $B$ we can write the action
(\ref{HLF11}) into a form that is linear in $\tilde{R}$:
\be
S_{F(\tilde{R})} = \int\rd^4 x \sqrt{\g} N \left[ B(\tilde{R} - A) + F(A) 
\right] \, .
\label{action_aux}
\ee
Variation with respect to $B$ yields $\tilde R = A$ that can be
inserted back into the action (\ref{action_aux}) in order to produce
the original action (\ref{HLF11}). The variation with respect to $A$
yields $B = F'(A)$.

First we rewrite $\tilde R$ in (\ref{action_aux}) into a more
explicit and useful form (see (\ref{HLF11}) for the definition
of $\tilde R$). The unit normal $n^\mu$ to the hypersurface
$\Sigma_t$ in space-time can be written in terms of the lapse
and the shift vector as $n^\mu = (n^0, n^i) =
\left(\frac{1}{N}, - \frac{N^i}{N}\right)$. The corresponding
one-form is $n_\mu = -N\nabla_\mu t = (-N, 0, 0, 0)$.
The term in (\ref{HLF11}) that involves the unit normal can be written
\be
\nabla_\mu \left( n^\mu \nabla_\nu n^\nu - n^\nu \nabla_\nu n^\mu \right)
= \nabla_\mu \left( n^\mu K \right) - \frac{1}{N} \g[ij] \nabla^{(3)}_i
\nabla^{(3)}_j N \, .
\ee
Thus we can rewrite $\tilde{R}$ as
\be
\tilde R = K_{ij} \mc{G}^{ijkl} K_{kl} + 2\mu  \nabla_\mu
\left( n^\mu K \right) - \frac{2\mu}{N} \g[ij] \nabla^{(3)}_i \nabla^{(3)}_j N
  - E^{ij} \mc{G}_{ijkl} E^{kl}\ .\label{tildeR_2nd}
\ee
Introducing (\ref{tildeR_2nd}) into (\ref{action_aux}) and
performing integrations by parts yields the action
\bea
S_{F(\tilde{R})} &=& \int\rd t \rd^3 \vect{x} \sqrt{\g}
\Bigl\{ N \left[ B \left( K_{ij} \mc{G}^{ijkl} K_{kl} - E^{ij}
\mc{G}_{ijkl} E^{kl} - A \right) + F(A) \right] \nn
&&\qquad\qquad\qquad \left.  - 2\mu K \left(  \dot{B} - N^i \p_i B \right)
  - 2\mu N \g[ij] \nabla^{(3)}_i \nabla^{(3)}_j B \right\} \, ,
\label{action_aux_final}
\eea
where the integral is taken over the union $\mc{U}$ of the
$t=\text{constant}$ hypersurfaces $\Sigma_t$ with $t$ over
some interval in $\mathbb{R}$, and we have written
$N n^\mu \nabla_\mu B = \dot{B} - N^i \p_i B$.
We assume that the boundary integrals on $\p\mc{U}$ and
$\p\Sigma_t$ vanish.

In the Hamiltonian formalism the field variables
$g_{ij}$, $N$, $N^i$, $A$ and $B$ have the canonically conjugated
momenta $\pi^{ij}$, $\pi_N$, $\pi_i$, $\pi_A$ and $\pi_B$,
respectively. For the spatial metric and the field $B$ we have
the momenta
\bea
\pi^{ij} &=& \frac{\delta S_{F(\tilde R)}}{\delta \dot{g}_{ij}}
= \sqrt{\g} \left[ B \mc{G}^{ijkl} K_{kl} - \frac{\mu}{N} \g[ij]
\left(  \dot{B} - N^i \p_i B \right) \right]\, ,\label{metric_momenta}\\
\pi_B &=& \frac{\delta S_{F(\tilde R)}}{\delta \dot{B}}
= - 2\mu \sqrt{\g} K \, .\label{pi_B}
\eea
We assume $\mu\neq 0$ so that the momentum (\ref{pi_B}) does not vanish.
Because the action does not depend on the time derivative of $N$,
$N^i$ or $A$, the rest of the momenta form the set of primary
constraints:
\be
\pi_N \approx 0\, ,\quad \pi_i(\vect{x}) \approx 0\, ,
\quad \pi_A(\vect{x}) \approx 0\, .
\label{p_constraints}
\ee
We consider $N$ to be projectable, i.e. $N = N(t)$, and therefore
also the momentum $\pi_N = \pi_N(t)$ is constant on $\Sigma_t$
for each $t$. The Poisson brackets are postulated in the form
(equal time $t$ is understood)
\bea
&& \{ \g_{ij}(\vect{x}),
\pi^{kl}(\vect{y}) \} = \frac{1}{2} \left( \delta_i^k \delta_j^l +
\delta_i^l \delta_j^k \right) \delta(\vect{x} - \vect{y})\, ,\nn
&& \{ N, \pi_N \} = 1\, ,\quad \{ N^i(\vect{x}), \pi_j(\vect{y}) \} =
\delta^i_j \delta(\vect{x} - \vect{y})\, ,\nn && \{ A(\vect{x}),
\pi_A(\vect{y}) \} = \delta(\vect{x} - \vect{y})\, ,\quad \{
B(\vect{x}), \pi_B(\vect{y}) \} = \delta(\vect{x} - \vect{y})\, ,
\eea
with all the other Poisson brackets vanishing. We shall continue
to omit the argument $(\vect{x})$ of the fields when there is
no risk of confusion. In order to obtain the Hamiltonian, we first
solve (\ref{metric_momenta})--(\ref{pi_B}) for $K_{ij}$ and $\dot{B}$,
\bea
K_{ij} &=& \frac{1}{\sqrt{\g}} \left[ \frac{1}{B} \left( \g_{ik} \g_{jl} 
\pi^{kl} - \frac{1}{3}
\g_{ij} \g_{kl} \pi^{kl} \right) - \frac{1}{6\mu} \g_{ij} \pi_B \right] \, ,\nn
\dot{B} &=& N^i \p_i B - \frac{N}{3\mu\sqrt{\g}} \left( \g_{ij} \pi^{ij}
+ \frac{1-3\lambda}{2\mu} B \pi_B \right) \, ,\label{dotB}
\eea
and further obtain $\dg_{ij} = 2NK_{ij} + \nabla^{(3)}_i N_j + \nabla^{(3)}_j 
N_i$.
Therefore both $\g_{ij}$ and $B$ are dynamical variables and
no more primary constraints are needed. The Hamiltonian is then
defined
\be
H = \int \rd^3 \vect{x} \left( \pi^{ij} \dg_{ij} + \pi_B \dot{B} \right) - L
= \int \rd^3 \vect{x} \left( N \mc{H}_0 + N^i \mc{H}_i \right)\, ,\label{Ha}
\ee
where the Lagrangian $L$ is defined by the action (\ref{action_aux_final}),
$S_{F(\tilde{R})} = \int\rd t L$, and the so called Hamiltonian
constraint and the momentum constraint are found to be
\bea
\mc{H}_0 &=& \frac{1}{\sqrt{\g}} \left[ \frac{1}{B} \left( \g_{ik} \g_{jl}
   \pi^{ij} \pi^{kl}
  - \frac{1}{3}\left( \g_{ij} \pi^{ij} \right)^2 \right)
  - \frac{1}{3\mu} \g_{ij} \pi^{ij} \pi_B
  - \frac{1-3\lambda}{12\mu^2} B \pi_B^2 \right] \nn
&& +
\sqrt{\g} \left[ B \left( E^{ij} \mc{G}_{ijkl} E^{kl} + A \right) - F(A)
+ 2\mu \g[ij] \nabla^{(3)}_i \nabla^{(3)}_j B \right] \, ,\nn
\mc{H}_i &=& - 2\g_{ij}\nabla^{(3)}_k \pi^{jk} + \nabla^{(3)}_i B \pi_B \nn
&=& -2\g_{ij}\p_k \pi^{jk} - \left( 2\p_j \g_{ik} - \p_i
\g_{jk} \right) \pi^{jk} + \p_i B \pi_B \, ,\label{Hb}
\eea
respectively. Again we assume that the boundary term resulting
from an integration by parts vanishes. We define the total
Hamiltonian by
\be
H_T = H + \lambda_N \pi_N + \int \rd^3 \vect{x} \left( \lambda^i \pi_i
+ \lambda_A \pi_A \right)\, ,
\label{H_T}
\ee
where the primary constraints (\ref{p_constraints}) are multiplied
by the Lagrange multipliers $\lambda_N$, $\lambda^i$, $\lambda_A$.
Note that there is no space integral over the product $\lambda_N \pi_N$
since they depend only on the time coordinate $t$ due to the
projectability of $N$.

The primary constraints (\ref{p_constraints}) have to be preserved
under time evolution of the system:
\bea
\dot{\pi}_N &=& \{ \pi_N, H_T \} = - \int
\rd^3 \vect{x} \mc{H}_0 \, ,\nn
\dot{\pi}_i &=& \{ \pi_i, H_T \} = - \mc{H}_i \, ,\nn
\dot{\pi}_A &=& \{ \pi_A, H_T \} = \sqrt{g^{(3)}} N
\left( - B + F'(A) \right)\, .
\eea
Therefore we impose the secondary constraints:
\bea
\Phi_0 &\equiv& \int \rd^3 \vect{x} \mc{H}_0 \approx 0 \, ,\nn
\Phi_i(\vect{x}) &\equiv& \mc{H}_i(\vect{x}) \approx 0 \, ,\nn
\Phi_A(\vect{x}) &\equiv& B(\vect{x}) - F'(A(\vect{x}))
\approx 0 \, .\label{s_constraints}
\eea
Here the Hamiltonian constraint $\Phi_0$ is global and the other two,
the momentum constraint $\Phi_i(\vect{x})$ and the constraint
$\Phi_A(\vect{x})$, are local. It is convenient to introduce a
globalized version of the momentum constraints $\Phi_i$:
\be
\Phi_S(\xi^i) \equiv \int \rd^3 \vect{x}\xi^i \mc{H}_i \approx 0 \, ,
\ee
where $\xi^i, i=1,2,3$ are three arbitrary smearing functions ---
the choices $\xi^i = \delta^i_j \delta(\vect{x}-\vect{y})$ will
produce the three local constraints $\mc{H}_j$ which in turn
imply the smeared one.

The total Hamiltonian (\ref{H_T}) can be written in
terms of the constraints as
\be
H_T = N\Phi_0 + \Phi_S(N^i) + \lambda_N \pi_N
+ \int \rd^3 \vect{x} \left( \lambda^i \pi_i
+ \lambda_A \pi_A \right)\, .
\label{H_T_as_constraints}
\ee

The consistency of the system requires that also the secondary
constraints $\Phi_0$, $\Phi_S(\xi^i)$ and $\Phi_A(\vect{x})$ have
  to be preserved under time evolution:
\bea
\dot{\Phi}_0 &=& \{ \Phi_0, H_T \} = N\{ \Phi_0, \Phi_0 \}
+ \{ \Phi_0, \Phi_S(N^i) \} + \int\rd^3 \vect{x} \lambda_A(\vect{x})
\{ \Phi_0, \pi_A(\vect{x}) \} \approx 0\, ,\nn
\dot{\Phi}_S(\xi^i) &=& \{ \Phi_S(\xi^i), H_T \}
= N \{ \Phi_S(\xi^i), \Phi_0 \} + \{ \Phi_S(\xi^i), \Phi_S(N^i) \} \approx 0\nn
\dot{\Phi}_A(\vect{x}) &=& \{ \Phi_A(\vect{x}), H_T \}
= N \{ \Phi_A(\vect{x}), \Phi_0 \} + \{ \Phi_A(\vect{x}), \Phi_S(N^i) \}
+ \int\rd^3 \vect{y} \lambda_A(\vect{y}) \{ \Phi_A(\vect{x}), \pi_A(\vect{y})
\}
\approx 0 \, ,\label{s_constraints_consistency}
\eea
where we have used the fact that the constraints $\pi_N$ and
$\pi_i$ have strongly vanishing Poisson brackets with every constraint.
We need to calculate the rest of the algebra of the constraints
under the Poisson bracket. The Poisson brackets between the
constraint $\Phi_S(\xi^i)$ and the canonical variables are
\bea
\{ \Phi_S(\xi^i), B \} &=& - \xi^i \p_i B\, ,\nn
\{ \Phi_S(\xi^i), \pi_B \} &=& - \p_i \left( \xi^i \pi_B \right)\, ,\nn
\{ \Phi_S(\xi^k), \g_{ij} \} &=& - \xi^k \p_k \g_{ij} - \g_{ik} \p_j
\xi^k - \g_{jk} \p_i \xi^k\, ,\nn
\{ \Phi_S(\xi^k), \pi^{ij} \} &=& - \p_k \left( \xi^k \pi^{ij} \right)
+ \pi^{ik} \p_k \xi^j +\pi^{jk} \p_k \xi^i\, ,\label{diffeom_generator}
\eea
and trivially zero for $A$ and $\pi_A$,
\be
\{ \Phi_S(\xi^i), A \} = 0\, ,\quad \{ \Phi_S(\xi^i), \pi_A \} = 0 \, .
\label{diffeom_generator2}
\ee
Thus $\Phi_S(\xi^i)$ generates the spatial diffeomorphisms for
the variables $B, \pi_B, \g_{ij}, \pi^{ij}$, and consequently
for any function or functional constructed from these variables,
and treates the variables $A, \pi_A$ as constants. By using this
result (\ref{diffeom_generator})--(\ref{diffeom_generator2})
we obtain the Poisson brackets for the constraints $\Phi_0$
and $\Phi_S(\xi^i)$:
\be
\{ \Phi_0, \Phi_0 \} = 0 \, ,\quad \{ \Phi_S(\xi^i), \Phi_0 \} = 0 \, ,
\quad \{ \Phi_S(\xi^i), \Phi_S(\eta^i) \} = \Phi_S(\xi^j \p_j \eta^i
  -  \eta^j \p_j \xi^i) \approx 0 \, .\label{Phi0_PhiS_PBs}
\ee
For the constraints $\pi_A$ and $\Phi_A(\vect{x})$ the Poisson
brackets that do not vanishing strongly are:
\bea
&&\{ \pi_A(\vect{x}), \Phi_0 \} = - \sqrt{\g} \Phi_A(\vect{x})
\approx 0 \, ,\quad \{ \pi_A(\vect{x}), \Phi_A(\vect{y}) \}
= F''(A(\vect{x})) \delta(\vect{x}-\vect{y})\nn
&&\{ \Phi_0, \Phi_A(\vect{x}) \} = \frac{1}{3\mu\sqrt{\g}}
\left(\g_{ij} \pi^{ij} + \frac{1-3\lambda}{2\mu}B\pi_B \right) \, ,
\quad \{ \Phi_S(\xi^i), \Phi_A(\vect{x}) \} = - \xi^i \p_i B \, .
\label{piA_PhiA_PBs}
\eea
Thus, in order to satisfy the consistency conditions
  (\ref{s_constraints_consistency}), we have to impose the tertiary constraint
\be
\Phi_\mr{ter} \equiv N^i \p_i B - \frac{N}{3\mu\sqrt{\g}}
\left(\g_{ij} \pi^{ij} + \frac{1-3\lambda}{2\mu}B\pi_B \right)
  - \lambda_A F''(A) \approx 0\, .\label{t_constraint}
\ee
Since $F''(A)=0$ would essentially reproduce the original projectable
Ho\v{r}ava-Lifshitz gravity, we assume that $F''(A) \neq 0$.
The first two terms in (\ref{t_constraint}), i.e. the expression
for $\dot{B}$ in (\ref{dotB}), does not vanish due to the
established constraints (\ref{p_constraints}) and (\ref{s_constraints}).
Therefore (\ref{t_constraint}) is a restriction on the  Lagrange
multiplier $\lambda_A$, and we can solve it from $\Phi_\mr{ter}=0$:
\be
\lambda_A = \frac{1}{F''(A)} \left( N^i \p_i B - \frac{N}{3\mu\sqrt{\g}}
\left(\g_{ij} \pi^{ij} + \frac{1-3\lambda}{2\mu}B\pi_B \right) \right)\, 
.\label{lambda_A}
\ee
Introducing (\ref{lambda_A}) into the Hamiltonian
(\ref{H_T_as_constraints}) ensures that now all the constraints
of the system are consistent.

According to the Poisson brackets (\ref{Phi0_PhiS_PBs})--(\ref{piA_PhiA_PBs})
between the constraints, we can set the second-class constraints
$\pi_A(\vect{x})$ and $\Phi_A(\vect{x})$ to vanish strongly, and
as a result turn the Hamiltonian constraint $\Phi_0$ and the
momentum constraint $\Phi_S(\xi^i)$ into first-class constraints.
For this end, we replace the Poisson bracket with the Dirac bracket,
which is given by
\be \label{DB}
\{ f(\vect{x}), h(\vect{y}) \}_\mr{DB} = \{ f(\vect{x}), h(\vect{y}) \}
+ \int\rd^3 \vect{z} \frac{1}{F''(A(\vect{z}))} \left( \{ f(\vect{x}),
\pi_A(\vect{z}) \} \{ \Phi_A(\vect{z}), h(\vect{y}) \} - \{ f(\vect{x}),
\Phi_A(\vect{z}) \} \{ \pi_A(\vect{z}), h(\vect{y}) \} \right) \, ,
\ee
where $f$ and $h$ are any functions of the canonical variables.
Assuming we can solve the constraint $\Phi_A(\vect{x})=0$,
i.e. $B=F'(A)$, for $A=\tilde{A}(B)$, where $\tilde{A}$ is the
inverse of the function $F'$, we can eliminate the variables $A$
and $\pi_A$. Thus the final variables of the system are
$\g_{ij}, \pi^{ij}, B, \pi_B$. The lapse $N$ and the shift vector
$N^i$, together with $\lambda_N$ and $\lambda^i$, are non-dynamical
multipliers. Then since every dynamical variable has a vanishing
Poisson bracket with the constraint $\pi_A$, the Dirac bracket
(\ref{DB}) reduces to the Poisson bracket,
\be \label{DB2}
\{ f(\vect{x}), h(\vect{y}) \}_\mr{DB} = \{ f(\vect{x}), h(\vect{y}) \}\, .
\ee

Finally the total Hamiltonian is the sum of the first-class constraints
\be
H_T = N\Phi_0 + \Phi_S(N^i) + \lambda_N \pi_N + \int\rd^3 \vect{x}
\lambda^i \pi_i \, .\label{H_T_sum_of_first-class}
\ee
It defines the equations of motion for every function $f(\vect{x})$
(or functional $f$) of the canonical variables
\be
\dot{f}(\vect{x}) = \{ f(\vect{x}), H_T \}  = N \{ f(\vect{x}), \Phi_0 \}
+ \{ f(\vect{x}), \Phi_S(N^i) \} + \lambda_N \{ f(\vect{x}), \pi_N \}
+ \int \rd^3 \vect{y} \lambda^i(\vect{y}) \{ f(\vect{x}), \pi_i(\vect{y}) \}\, 
.
\label{EOM_general}
\ee

We have calculated the Hamitonian (\ref{Ha})--(\ref{Hb}) of
the proposed modified Ho\v{r}ava-Lifshitz $F(R)$-gravity and
established the preservation of the primary constraints
(\ref{p_constraints}) by imposing the required secondary
constraints (\ref{s_constraints}), including the Hamiltonian
constraint and the momentum constraint. In order to ensure
the consistency of the secondary constraints we introduced
the tertiary constraint (\ref{t_constraint}) that was used to
fix the Lagrange multiplier $\lambda_A$ of the primary constraint
$\pi_A$. Finally, we eliminated the pair of variables $A, \pi_A$ by
imposing the second-class constraints $\pi_A$ and $\Phi_A$, and
introduced the Dirac bracket (\ref{DB}) that reduced to (\ref{DB2}).
The total Hamiltonian was obtained in its final form
(\ref{H_T_sum_of_first-class}) as a sum of the first-class constraints.
We conclude that the proposed action (\ref{HLF11}) of this
modified $F(R)$ Ho\v{r}ava-Lifshitz gravity, which obeys the
projectability condition, defines a consistent theory.
This conclusion agrees with the recent analysis of our theory 
presented in ref.~\cite{Kluson:2010xx}.

\section{FRW cosmology for some versions of modified Ho\v{r}ava-Lifshitz 
$F(R)$-gravity.}

This section is devoted to the study of the FRW Eqs. (\ref{HLF13})
and (\ref{HLF14}) which admit a de Sitter universe solution. We now
neglect the matter contribution by putting $p=\rho=0$. Then by
assuming $H=H_0$, both of Eq. (\ref{HLF13}) and (\ref{HLF14}) lead
to the same equation
\be
\label{HLF18}
0 = F\left( 3 \left(1 - 3 \lambda + 6 \mu \right) H_0^2 \right)
  - 6 \left(1 - 3\lambda + 3\mu\right) H_0^2 F'\left( 3 \left(1 - 3
\lambda + 6 \mu \right) H_0^2 \right) \, ,
\ee
as long as the integration constant vanishes ($C=0$) in Eq. (\ref{HLF16}).

First we consider the popular case that
\be \label{HLF19}
F\left(\tilde R\right) \propto \tilde R + \beta \tilde R^2 \, .
\ee
Then Eq. (\ref{HLF18}) gives
\be
\label{HLF20}
0 = H_0^2 \left\{ 1 -
3\lambda + 9\beta \left(1 - 3\lambda + 6 \mu \right) \left( 1 -
3\lambda + 2\mu \right) H_0^2 \right\}\, .
\ee
In the case of usual
$F(R)$-gravity, where $\lambda=\mu=1$ and therefore $1 - 3\lambda +
2\mu =0$, there is only the trivial solution $H_0^2 = 0$, although
the $R^2$-term could generate the inflation when more gravitational
terms, like $R_{\mu\nu}R^{\mu\nu}$ etc., are added. For our general
case, however, there exists the non-trivial solution
\be
\label{HLF21}
H_0^2 = - \frac{ 1 - 3\lambda }{\beta \left(1 -
3\lambda + 6 \mu \right) \left( 1 - 3\lambda + 2\mu \right)}\, ,
\ee
as long as the r.h.s. of (\ref{HLF21}) is positive. If the magnitude
of this non-trivial solution is small enough, this solution might
correspond to the accelerating expansion in the present universe.
Hence, the $R^2$-term may generate the late-time acceleration. On
the other hand, the above solution may serve as an inflationary
solution for the early universe (with the corresponding choice of
parameters).

Instead of (\ref{HLF19}) one may consider the following model:
\be
\label{HLF22}
F\left(\tilde R\right) \propto \tilde R + \beta \tilde
R^2 + \gamma \tilde R^3\, .
\ee
Then Eq. (\ref{HLF18}) becomes
\be
\label{HLF23}
0 = H_0^2 \left\{ 1 - 3\lambda + 9\beta \left(1 -
3\lambda + 6 \mu \right) \left( 1 - 3\lambda + 2\mu \right) H_0^2 +
9\gamma \left(1 - 3\lambda + 6 \mu \right)^2 \left( 5 - 15\lambda +
12 \mu \right) H_0^4 \right\}\, ,
\ee
which has the following two non-trivial solutions,
\be
\label{HLF24} H_0^2 = - \frac{ \left( 1 -
3\lambda + 2\mu \right) \beta }{2 \left(1 - 3\lambda + 6 \mu \right)
\left( 5 - 15\lambda + 12 \mu \right) \gamma} \left( 1 \pm \sqrt{ 1
  - \frac{4 \left(1 - 3\lambda \right)\left( 5 - 15\lambda + 12 \mu
\right) \gamma} { 9 \left( 1 - 3\lambda + 2\mu \right)^2 \beta^2} }
\right)\, ,
\ee
as long as the r.h.s. is real and positive. If
\be
\label{HLF25}
\left| \frac{4 \left(1 - 3\lambda \right)\left( 5 -
15\lambda + 12 \mu \right) \gamma} { 9 \left( 1 - 3\lambda + 2\mu
\right)^2 \beta^2} \right| \ll 1\, ,
\ee
one of the two solutions is
much smaller than the other solution. Then one may regard that the
larger solution corresponds to the inflation in the early universe
and the smaller one to the late-time acceleration, similarly to the
modified gravity model \cite{Nojiri:2003ft}, where such unification
has been first proposed. The fact that such two solutions are 
connected could be demonstrated by numerical calculation. 
Note that some of the above models may possess the future singularity 
in the same way as the usual $F(R)$-gravity. However, it would be 
possible to demonstrate that adding terms wtih even higher derivatives 
might cure this singularity, similarly as the addition of the 
$R^2$-term did in the usual $F(R)$-gravity. 
Hence, we have suggested the qualitative possibility to unify 
the early-time inflation with the late-time acceleration in the 
modified Ho\v{r}ava-Lifshitz $F(R)$-gravity.

\section{More general action}\label{gen_action}

In the formulation of $F(R)$ Ho\v{r}ava-Lifshitz-like gravity, we do
not require full diffeomorphism-invariance,  but only invariance
under ``foliation-preserving'' diffeomorphisms (\ref{fpd1}).
Therefore there are many invariants or covariant quantities made
from the metric like $K$, $K_{ij}$, $\nabla^{(3)}_i K_{jk}$,
$\cdots$, $\nabla^{(3)}_{i_1} \nabla^{(3)}_{i_2} \cdots
\nabla^{(3)}_{i_n} K_{jk}$, $R^{(3)}$, $R^{(3)}_{ij}$,
$R^{(3)}_{ijkl}$, $\nabla^{(3)}_i R^{(3)}_{jk}$, $\cdots$, $\nabla_\mu
\left( n^\mu \nabla_\nu n^\nu - n^\nu \nabla_\nu n^\mu \right)$,
$\cdots$, etc. Then the action composed of such invariants as
\bea
\label{HLF26}
S_\mathrm{gHL} &=& \int \rd^4 x \sqrt{g^{(3)}} N F
\left(g^{(3)}_{ij}, K, K_{ij}, \nabla^{(3)}_i K_{jk}, \cdots,
\nabla^{(3)}_{i_1} \nabla^{(3)}_{i_2} \cdots \nabla^{(3)}_{i_n}
K_{jk}, \right. \nn
&& \left. \cdots, R^{(3)}, R^{(3)}_{ij},
R^{(3)}_{ijkl}, \nabla^{(3)}_i R^{(3)}_{jk}, \cdots, \nabla_\mu \left(
n^\mu \nabla_\nu n^\nu - n^\nu \nabla_\nu n^\mu \right) \right)\, ,
\eea
could be a rather general action for the generalized
Ho\v{r}ava-Lifshitz gravity. Note that one can also include the
(cosmological) constant in the above action. Here it has been
assumed that the action does not contain derivatives higher than the
second order with respect to the time coordinate $t$. In the usual
$F(R)$-gravity, there appears the extra scalar mode since the
equations given by the variation over the metric tensor contain the
fourth derivative. Now we avoid such extra modes except the one
scalar mode.

In the FRW space-time (\ref{HLF8}) with the flat spatial part and
non-trivial $N=N(t)$, we find
\bea
\label{HLF27}
&& \Gamma^0_{00} =
\frac{\dot N}{N}\, , \quad \Gamma^0_{ij} = \frac{a^2 H}{N^2}\delta_{ij}\, ,
\quad \Gamma^i_{j0} = H\delta^i_{\ j}\, \quad
\mbox{other}\ \Gamma^\mu_{\nu\rho} = 0\, , \nn
&& K_{ij} =
\frac{a^2H}{N}\delta_{ij}\, ,\quad \nabla^{(3)}_i = 0\, , \quad
R^{(3)}_{ijkl}=0\, ,\quad \nabla_\mu \left( n^\mu \nabla_\nu n^\nu -
n^\nu \nabla_\nu n^\mu \right) = \frac{3}{a^3 N}\frac{\rd}{\rd
t}\left(\frac{a^3 H}{N}\right)\, .
\eea
Then one gets
\bea
\label{HLF27b}
&& g^{(3)}_{ij}=a^2\delta_{ij}\, , \quad K=\frac{3 H}{N}\, ,
\quad \nabla^{(3)}_i K_{jk}= \cdots = \nabla^{(3)}_{i_1}
\nabla^{(3)}_{i_2} \cdots \nabla^{(3)}_{i_n} K_{jk} = \cdots = 0 \, ,\nn
&& R^{(3)}=R^{(3)}_{ij}=R^{(3)}_{ijkl}=\nabla^{(3)}_i R^{(3)}_{jk}=
\cdots =0\, ,
\eea
and since $F$ must be a scalar under the spatial
rotation, the action (\ref{HLF26}) reduces to
\bea
\label{HLF28}
S_\mathrm{gHL} &=& \int \rd^4 x \sqrt{g^{(3)}} N F
\left(\frac{H}{N}, \frac{3}{a^3 N}\frac{\rd}{\rd t}\left(\frac{a^3
H}{N}\right)\right)\, .
\eea
Therefore, if we consider the FRW
cosmology, the function $F$ should depend on only two variables,
$\frac{H}{N}$ and $\frac{3}{a^3 N}\frac{\rd}{\rd t}\left(\frac{a^3
H}{N}\right)$. For instance, $\tilde R$ in (\ref{HLF12}) is given by
this combination. As an illustrative example, we may consider the
following one:
\be
\label{HLF29}
F = f_0 \left(K^{ij}K_{ij}
  - \lambda K^2 \right) + f_1 \nabla_\mu \left( n^\mu \nabla_\nu n^\nu
  - n^\nu \nabla_\nu n^\mu \right)^2 \, .
\ee
Then in the FRW space-time
(\ref{HLF2}), by the variation of the scale factor $a$, we obtain
the following equation:
\be
\label{HLF30}
0 = 2 f_0 \left( 1 -
3\lambda \right)\left(H^2 + \dot H\right) + 3 f_1 \left( 27 H^4
+ 54 H^2 \dot H + 15 {\dot H}^2 + 18 H \ddot H + 2 \dddot H \right)\, .
\ee
If we assume a de Sitter universe $H=H_0$ with constant $H_0$,
Eq. (\ref{HLF30}) reduces to
\be
\label{HLF31}
0 = 2 f_0 \left( 1 - 3\lambda \right) H^2 + 81 f_1 H^4 \, ,
\ee
which has the non-trivial solution
\be
\label{HLF32}
H^2 = - \frac{2 f_0 \left( 1 - 3\lambda \right)}{81 f_1}\, ,
\ee
as long as the r.h.s. is positive. In the
same way, a large class of modified Ho\v{r}ava-Lifshitz gravities
may be constructed. For instance, one can construct
Ho\v{r}ava-Lifshitz-like generalizations of $F(G)$-gravity where
the action is the Einstein-Hilbert term plus a function $F$ of
the Gauss-Bonnet invariant $G$,
non-local gravity, $F(R, R_{\mu\nu}R^{\mu\nu}, R_{\mu\nu\alpha\beta}
R^{\mu\nu\alpha\beta})$, etc. It is remarkable that some special
subclass of such Ho\v{r}ava-Lifshitz-like theories will have the
same spatially-flat FRW background dynamics as the corresponding
traditional modified gravity.

\section{Discussion}

We have suggested a quite general approach for the modification of
Ho\v{r}ava-Lifshitz gravity. We concentrated mainly on the
$F(R)$-gravity version. The consistency of its spatially-flat FRW
field equations has been demonstrated. 
The Hamiltonian and the corresponding constraints of the modified 
$F(R)$ Ho\v{r}ava-Lifshitz gravity have been derived. It has been shown 
that these constraints are consistent under the dynamics of the system, 
and that they do not constrain the physical degrees of freedom too 
much. It is demonstrated that a
degenerate subclass of the proposed general modified $F(R)$
Ho\v{r}ava-Lifshitz gravity corresponds to the earlier proposed
$F(R)$ extension of Ho\v{r}ava-Lifshitz gravity. The preliminary
study of FRW cosmology indicates a possibility to describe or even
to unify the early-time inflation with the late-time acceleration \cite{S}.
The motivation to consider such a theory is clear: it includes 
conventional $F(R)$-gravity and Ho\v{r}ava-Lifshitz gravity as 
limiting cases. The former offers interesting cosmological solutions, 
while the latter may hold the promise of UV-completeness.

Our proposal opens the bridge between the conventional modified
gravity and its Ho\v{r}ava-Lifshitz counterpart. Indeed, it is
demonstrated that our model with a special choice of parameters
($\lambda = \mu = 1$) leads to the same spatially-flat FRW dynamics
as its traditional counterpart, which is fully
diffeomorphism-invariant. Moreover, we eventually proposed the most
general construction for a modified gravity that is invariant under
foliation-preserving diffeomorphisms. In this way, any traditional
modified gravity has its counterpart, where the Lorentz symmetry is
broken. The explicit construction may be made using the results of
Section \ref{gen_action}. Having in mind that a number of
traditional modified theories of gravity are cosmologically viable
and pass the local tests, one can expect that it will eventually be
possible to realize any accelerating FRW cosmology in this modified
Ho\v{r}ava-Lifshitz theory. This will be studied elsewhere.

\section*{Acknowledgments}

This research has been supported in part by MEC (Spain) project
FIS2006-02842 and AGAUR(Catalonia) 2009SGR-994 (SDO), by Global COE
Program of Nagoya University (G07) provided by the Ministry of
Education, Culture, Sports, Science \& Technology (SN). M. O. is
supported by the Finnish Cultural Foundation. The support of the
Academy of Finland under the Projects No. 121720 and 127626 is
greatly acknowledged.

\end{document}